\documentclass[aps,prl,twocolumn,superscriptaddress,preprintnumbers,nofootinbib]{revtex4-1}
\usepackage{amsmath,amssymb,graphicx,natbib,aas_macros,bm}
\usepackage{enumerate}
\usepackage{hyperref}

\usepackage{color}

\begin{document} 

\preprint{EFI 12-24}
\preprint{ANL-HEP-PR-12-74}

\title{Higgs Couplings and Precision Electroweak Data}

\author{Brian Batell}
\email{batell@uchicago.edu}
\affiliation{Enrico Fermi Institute and Department of Physics, University of Chicago, Chicago, IL,
60637}

\author{Stefania Gori}
\email{goris@uchicago.edu}
\affiliation{Enrico Fermi Institute and Department of Physics, University of Chicago, Chicago, IL,
60637}
\affiliation{HEP Division, Argonne National Laboratory, 9700 Cass Ave., Argonne, IL 60439}

\author{Lian-Tao Wang}
\email{liantaow@uchicago.edu}
\affiliation{Enrico Fermi Institute and Department of Physics, University of Chicago, Chicago, IL,
60637}
\affiliation{Kavli Institute for Cosmological Physics, University of Chicago, Chicago, IL, 60637}

\begin{abstract}
In light of the discovery of a Higgs-like particle at the LHC, we revisit the status of the precision electroweak data, focusing on two discrepant observables: 1) the long-standing $2.4 \sigma$ deviation in the forward-backward asymmetry of the bottom quark $A^b_{FB}$, and 2) the $2.3 \sigma$ deviation in $R_b$, the ratio of the $Z \rightarrow b \bar b$ partial width to the inclusive hadronic width, which is now in tension after a recent calculation including new two-loop electroweak corrections.  
We consider possible resolutions of these discrepancies. Taking the data at face value, the most compelling scenario is that new physics directly affects $A_{FB}^b$ and $R_b$, bringing the prediction into accord with the measured values. We propose a modified `Beautiful Mirrors' scenario which contains new vector-like quarks that mix with the $b$ quark, modifying the $Z b\bar b$ vertex and thus correcting  $A_{FB}^b$ and $R_b$. 
We show that this scenario can lead to modifications to the production rates of the Higgs boson in certain channels, and in particular a sizable enhancement in the diphoton channel. We also describe additional collider tests of this scenario.
\end{abstract}

\maketitle

The ATLAS and CMS experiments have recently discovered a new boson with properties closely resembling those of the Standard Model (SM) Higgs boson~\cite{ATLAS-Higgs, CMS-Higgs}. The focus of the experimental collaborations now turns to determining the properties of this state, such as its quantum numbers and its couplings to other SM particles. This is an extremely important task since any deviation from the SM Higgs boson predictions will unambiguously point to new physics (NP) at the TeV scale. In this respect it is intriguing that both experiments observe a slight enhancement in the $h\rightarrow \gamma \gamma$ channel, though with the current dataset this enhancement is not statistically significant~\cite{ATLASgaga,CMSgaga}.

If, as the current data suggest, this new state is indeed the Higgs boson, its mass $m_h \sim 125$ GeV will be in accord with the expectation indirectly suggested by the precision electroweak data, which favors a light SM-like Higgs boson. 
Of the many precision measurements used to test the electroweak sector, most of them are in  good agreement with the SM predictions, 
see for example~\cite{LEPEWWG},\cite{Beringer:1900zz},\cite{Baak:2012kk}. 
However, there are a couple of notable discrepancies. The forward-backward asymmetry of the bottom quark $A_{FB}^b$ measured at the $Z$-pole at LEP1 constitutes a long-standing exception. The measured value and SM prediction are~\cite{LEPEWWG},~\cite{Baak:2012kk},
\begin{eqnarray}
(A_{FB}^b)_{\rm exp} &  =  & 0.0992 \pm 0.0016, \\
(A_{FB}^b)_{\rm SM}  & =  &  0.1032^{+0.0004}_{-0.0006} ,
\end{eqnarray}
and thus exhibit a 2.4$\sigma$ discrepancy. 
Furthermore, a recent calculation of $R_b$, the ratio of the $Z\rightarrow b \bar b$ partial width to the inclusive hadronic width, which includes new two-loop electroweak corrections, now puts the prediction in tension with the measured value~\cite{Freitas:2012sy}. 
The measurement and prediction read~\cite{LEPEWWG},~\cite{Baak:2012kk},
\begin{eqnarray}
(R_b)_{\rm exp} &  =  & 0.21629 \pm 0.00066, \\
(R_b)_{\rm SM}  & =  &0.21474\pm0.00003, 
\end{eqnarray}
displaying a 2.3$\sigma$ discrepancy.

It is a matter of debate whether these deviations call for NP. Among a large ensemble of measurements, such as that used in the electroweak fit, one may expect an occasional discrepancy of this size, which could be the result of a statistical fluctuation or unknown systematic error. However, a possible counterpoint with regards to $A_{FB}^b$, as emphasized in Refs.~\cite{Chanowitz:2001bv,Chanowitz:2002cd}, is that $A_{FB}^b$ favors a {\it heavy} Higgs. Excluding the measurement from the global fit would favor a Higgs boson much lighter than the LEP bound, perhaps suggesting NP that improves the electroweak fit with a 125 GeV Higgs. This approach was also advocated in Ref.~\cite{Altarelli:2001wx}. This discrepancy and its interplay with the indirect determination of the Higgs mass has stimulated a number of theoretical works that attempt to resolve this mystery, see {\it e.g.}~\cite{Erler:1996ww},\cite{Erler:1999nx},\cite{Altarelli:2001wx},\cite{Choudhury:2001hs},\cite{He:2002ha},\cite{Morrissey:2003sc},\cite{Kumar:2010vx},\cite{delAguila:2010mx},\cite{DaRold:2010as},\cite{Alvarez:2010js},\cite{Dermisek:2011xu}, \cite{Djouadi:2011aj},\cite{Dermisek:2012qx}.

With a direct and precise measurement of the Higgs mass from the LHC experiments, as well as the relatively recent tension in $R_b$, it is useful to revisit these arguments. At a very simplistic level, we observe that there are now two measurements related to bottom quarks which are discrepant, suggesting either a systematic effect associated with $b$-quarks or possibly NP in the $b$-quark sector. Thus, we will consider two possible resolutions to the  $A_{FB}^b$ and $R_b$ discrepancies: First, we will entertain the possibility that NP alters the $Z b\bar b$ vertex, thus changing the predictions for $A_{FB}^b$ and $R_b$. As we will demonstrate, this scenario leads to a dramatic improvement in the global fit of the data compared to the SM. Second, we will hypothesize that the $A_{FB}^b$ and $R_b$ deviations are a result of an unknown systematic effect, and as such remove these observables from the EW fit. In this case, we find that the remaining observables are well described by the SM, and NP which contribute to oblique parameters $S$ and $T$~\cite{Peskin:1991sw} can only marginally improve the fit.

Thus, if one takes the $A_{FB}^b$ and $R_b$ measurements seriously, the most compelling interpretation is that NP directly affects $A_{FB}^b$ and $R_b$ so as to bring the measured and predicted values into agreement. As an example we will focus on the ``Beautiful Mirrors'' scenario of Ref.~\cite{Choudhury:2001hs}. In this class of models, new vector-like quarks mix with the bottom quark of the SM, thus modifying the $Z \bar b_R b_R$ coupling and in turn  $A_{FB}^b$. We then expect that the Higgs production rates as well as decay modes will deviate from their SM values, since the $h \bar b b$ coupling is altered and new exotic vector-like quarks contribute to the  loop induced processes $h \rightarrow gg$ and $h \rightarrow \gamma \gamma$. We will revisit minimal models of this type, as well as some extensions to exhibit the range of possibilities in Higgs phenomenology.

\section{ Interpreting the Precision Electroweak Data}

In this section we will examine interpretations of the precision electroweak data using global fits, particularly focusing on possible solutions to the $A_{FB}^b$ and $R_b$ discrepancies. Specifically, there are two distinct scenarios we wish to test:

\begin{itemize}

\item New physics affects the $Z \bar b  b$ vertex, thus modifying the $A_{FB}^b$ and $R_b$.

\item The discrepancies in $A_{FB}^b$ and $R_b$ are attributed to an unknown systematic error, and the observables are thus removed from the global fit. 

\end{itemize}
Before presenting our results for these two scenarios, let us briefly describe the ingredients going into our global fit of the precision electroweak data. Our fit follows closely that of the Gfitter group~\cite{Baak:2012kk}.  
We use the high energy data, including $Z$-pole data from LEP and SLD~\cite{LEPEWWG}, as well as measurements of the top\cite{Aaltonen:2012ra}, 
$W$~\cite{TevatronElectroweakWorkingGroup:2012gb}, and Higgs mass~\cite{ATLAS-Higgs},\cite{CMS-Higgs} and the hadronic contribution to the 
running fine structure constant~\cite{Davier:2010nc}. Explicitly, we consider the following inputs in the fit:
\begin{eqnarray}
&m_Z,~ \Gamma_Z,~\sigma_{\rm had}^0, ~R_\ell, ~R_c, ~R_b, & \nonumber \\
&~ A_{FB}^\ell,~A_\ell,~A_c,~A_b,~A_{FB}^c,~A_{FB}^b,~\sin^2 \theta_{\rm eff},~& \nonumber \\
&m_W,~\Gamma_W,~m_t,~\Delta\alpha^{(5)}_{\rm had},~m_h.&
\label{data}
\end{eqnarray}
The experimental values for these observables are taken from~\cite{Baak:2012kk}.  
The leptonic asymmetry parameter $A_\ell$ is an average of the LEP and SLD measurements, while the heavy flavor observables $A_c,~A_b,~A_{FB}^c,~A_{FB}^b,~R_c,~R_b $ are averaged from LEP and SLC\cite{LEPEWWG}. We also account for correlations in $Z$ lineshape as well as heavy flavor observables~\cite{LEPEWWG}.
The $W$ boson mass is taken from~\cite{TevatronElectroweakWorkingGroup:2012gb}, which includes the recent precise measurements from CDF and D0, as well as older measurements from LEP and the Tevatron experiments. For the Higgs mass, we combine the ATLAS ($126.0\pm0.4\,({\rm{stat}})\pm 0.4\,({\rm{syst}})$ GeV~\cite{ATLAS-Higgs}) and CMS ($125.3\pm0.4\,({\rm{stat}})\pm 0.5\,({\rm{syst}})$ GeV~\cite{CMS-Higgs}) results, assuming no correlation of the systematic uncertainties, to obtain $m_h = 125.7 \pm 0.4$ GeV.

For the SM theory predictions we allow the following parameters to vary in the fit:
\begin{equation}
m_h, m_Z, m_t, \Delta \alpha^{(5)}_{\rm had}, \alpha_s.
\label{fitpar}
\end{equation}
We work with the running electromagnetic coupling which depends on the $\Delta \alpha^{(5)}_{\rm had}$ in (\ref{fitpar}).  For the $W$ boson mass prediction we utilize the numerical parameterization in Eqs.~(6,7,9) of Ref.~\cite{Awramik:2003rn}, which includes the full two loop prediction plus certain important higher order corrections.
We employ the numerical parameterizations of the $Z $ boson partial widths in terms of the SM parameters (\ref{fitpar}) from Ref.~\cite{Cho:2011rk} (updated from earlier work in Refs.~\cite{Hagiwara:1994pw}). From here we construct $\Gamma_Z, \sigma^0_{\rm had}, R_{\ell}, R_{c}$. 
For $R_b$, we instead use the recent result of Ref.~\cite{Freitas:2012sy}, which contains new two-loop electroweak corrections. For the effective weak mixing angle, we use the parameterizations in Refs.~\cite{Awramik:2006uz,Freitas:2012sy}, which account for the latest two-loop electroweak corrections. From these parameterizations, we calculate the asymmetry parameters $A_f$ and and the forward-backward asymmetries $A_{FB}^f$.

Additionally, we will explore NP scenarios  with oblique corrections parameterized with the Peskin-Takeuchi parameters $S$ and $T$,  as well as tree-level shifts to the $Z \bar b b $ vertex. 
In the presence of oblique corrections the theory predictions for the observables are modified to have the form ${\cal O} = {\cal O}_{\rm SM} + c_S S + c_T T$, where the coefficients $c_i$ depend on the input parameters (\ref{data}).These shifts have been computed in {\it e.g.} Ref.~\cite{Burgess:1993vc}, and we employ the expressions from this reference. We note that for the models considered in this paper the contribution to the $U$ parameter is small, so we will not consider it in our fits here.  

We parameterize the shifts in the $Z \bar b b$ coupling as
\begin{equation}
{\cal L } = \frac{g}{c_W} Z_\mu \bar b \gamma^\mu [ (g_{Lb} + \delta g_{Lb}) P_L + (g_{Rb} + \delta g_{Rb}) P_R    ]b,
\label{eq:Zbb}
\end{equation} 
where $c_W \equiv \cos \theta_W$, with $\theta_W$ the weak mixing angle, $g_{Lb}$ and $g_{Rb}$ are the SM left and right couplings given by $g_{Lb} = 1/2+s_W^2/3 \simeq -0.42 $, and $g_{Rb} = s_W^2/3 \simeq -0.077$.
The shifts $\delta g_{Lb,Rb}$ represent a non-universal contribution due to NP. These shifts enter into the predictions for observables in (\ref{fitpar}) depending on the $Z \rightarrow \bar b  b$ partial width as well as the asymmetry parameters $A_{FB}^b$ and $A_b$. Similar fits to $Z\bar b b$ couplings have been performed in the past~\cite{Bamert:1996px},\cite{Haber:1999zh},\cite{Choudhury:2001hs},\cite{Kumar:2010vx}.

We will now use the results of global fits to the data to examine the two scenarios described at the beginning of this section, namely 1) NP in $Z \bar b b$ couplings affects $A_{FB}^b$ and $R_b$, and 2) $A_{FB}^b$ and $R_b$ are a result of an unknown systematic error. 

\paragraph{\bf New physics in $Z \bar b b$ couplings}
%\scalebox{0.5}{
\begin{table*}[t]\small
\begin{center}
\renewcommand{\arraystretch}{1.3}
 \begin{tabular}{ | c || c | c | c | c | c  || c |c| }
\hline
\hline
%\multicolumn{8}{|c|}{    } \\ \hline \hline
%%
 Fit & $m_h$ (GeV)   &
 Pull[$A_{FB}^b$] & Pull[$R_b$] & $\delta g_{Lb}$  & $\delta g_{Rb}$   & $\chi^2/{\rm d.o.f} $ & $p$-value   
  \\
 \hline \hline
SM (All data) & $125.7 \pm 0.4 $ & $2.6 $ & $ -2.4 $ &  - &  -  &  20.5/13   & 0.08
\\  
\hline
SM + $\delta g_{Lb,Rb}$ ($+$) (All data) & $125.7\pm0.4$ & 0.7 & 0.2 & $0.001\pm 0.001$ & $0.016\pm0.005$  & 9.6/11   & 0.57
\\  
\hline
SM + $\delta g_{Lb,Rb}$ ($-$) (All data) & $125.7\pm0.4$ & 0.8& 0.2 & $0.001\pm 0.001$ & $-0.170\pm0.005$  & 9.3/11   & 0.60
\\  
\hline
\hline
\end{tabular}
\end{center}
\caption{ {\it New physics in $Z \bar b b$}:  Results of the global fits to the precision electroweak observables in Eq.~(\ref{data}) are displayed for 1) the SM hypothesis  and 2) the hypothesis of NP in $Z \bar b b$ couplings as in Eq.~(\ref{eq:Zbb}). We show the Higgs mass prediction, the pull values for $A_{FB}^b$ and $R_b$, and the goodness-of-fit in terms of $\chi^2/{\rm d.o.f.}$ and $p$-value. For the second hypothesis, we also show the best-fit regions for the coupling shifts ($\delta g_{Lb}$, $\delta g_{Rb}$). Note that there are two best-fit regions regions (labeled by ``$+$" and ``$-$'').}
\label{tab:fit1}
\end{table*}
%}

Here we take seriously the measurements of $A_{FB}^b$ and $R_b$ and include them in the global fit along with the other observables in Eq.~(\ref{data}). Before considering NP, we would first like to understand how well the SM describes the data.  The results of our fit to the SM are presented in Table~\ref{tab:fit1}.  
For the best-fit point, we obtain $\chi^2/{\rm d.o.f.} = 20.5/13$ corresponding to a $p$-value is 0.08. The goodness-of-fit is marginal due to the discrepancies in $A_{FB}^b$ ($2.6\sigma$) and $R_b$ ($2.4\sigma$). The Higgs mass is nailed down by the LHC measurement, with a value of $m_h = 125.7 \pm 0.4$ GeV. 

\begin{figure}[h!]
\begin{center}
\includegraphics[width=.45\textwidth]{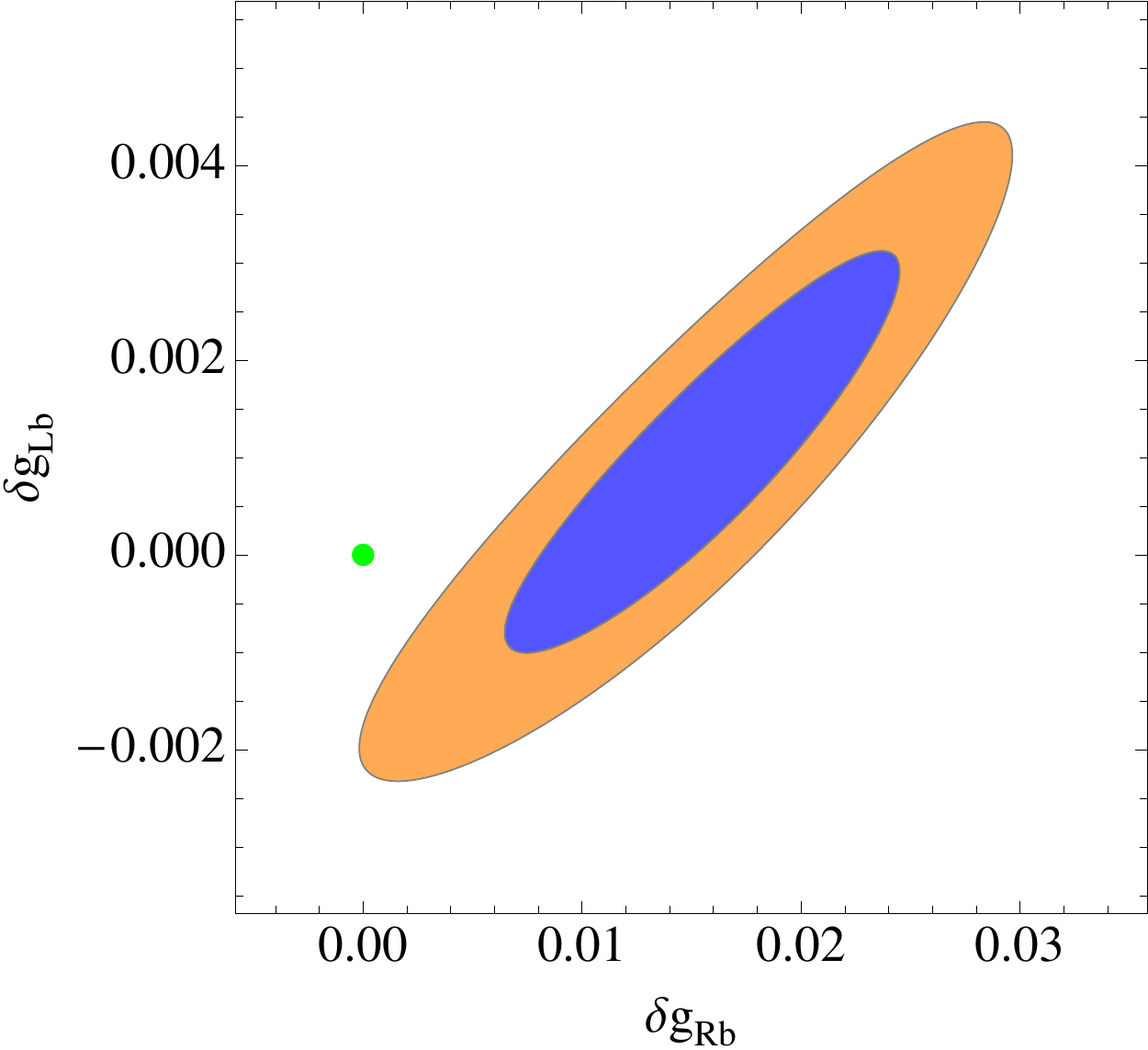}~
\caption{Best-fit region for the $Z \bar  b b$ coupling shifts $\delta g_{Lb}$, $\delta g_{Rb}$. We show here the region with a moderate positive shift to $\delta g_{Rb}$. The SM corresponds to the green point. }
\label{fig:dg}
\end{center}
\end{figure}

We now hypothesize the existence of NP that causes non-universal shifts in the $Z \bar b  b$ vertex as in Eq.~(\ref{eq:Zbb}). The results of the fit are displayed in Table~\ref{tab:fit1}. The quality of the fit is much improved compared to the SM fit. There are two preferred regions in the fit, one characterized by a large negative shift in the right coupling 
$\delta g_{Rb} = - 0.17\pm 0.05$ (labeled ``-'')  and the other characterized by a moderate positive shift of the right coupling $\delta g_{Rb} =0.016\pm 0.005$ (labeled ``+''). 
In both cases, the data also suggest a small positive shift in the left coupling $\delta g_{Lb} = 0.001\pm 0.001$. The improvement in the fit quality stems from the fact that the $A_{FB}^b$ and $R_b$ predictions are no longer discrepant, lying within $1\sigma$ of their measured values. In Fig.~(\ref{fig:dg}), we show the preferred region in the $\delta g_{Lb}$-$\delta g_{Rb}$ parameter space for the solution with a moderate positive $\delta g_{Rb}$ shift, which will be relevant below when we investigate explicit models of NP that cause shifts to the $Z \bar b b$ vertex.  

The asymmetry parameter $A_b$ was measured at SLD to be $(A_b)_{\rm exp} = 0.923\pm0.020$,  which is in agreement with our SM prediction  $(A_b)_{\rm SM} = 0.9346\pm0.0003$, is modified by the shifts $\delta g_{Lb,Rb}$. For example, at the best fit point ($\delta g_{Rb} = 0.016$, $\delta g_{Lb} = 0.001$), we predict $(A_b)_{\rm th} = 0.907\pm0.009$, still within $1\sigma$ of the measured value. The larger error in the theory prediction is mainly due to the uncertainty in $\delta g_{Rb}$.

\paragraph{\bf $A_{FB}^b$ and $R_b$ due to a systematic error }

%\scalebox{0.5}{
\begin{table*}[t]\footnotesize
\begin{center}
\renewcommand{\arraystretch}{1.2}
 \begin{tabular}{ | c || c | c | c | c | c  || c |c| }
\hline
\hline
%\multicolumn{8}{|c|}{    } \\ \hline \hline
%%
 Fit & $m_h$ (GeV)   &
 $S$ & $T$ & $\chi^2/{\rm d.o.f} $ & $p$-value   
  \\
 \hline \hline
\hline
SM (No $A_{FB}^b,R_b$) & $125.7\pm0.4$ & - & -  & 9.3/12   & 0.67
\\  
\hline
SM (No $A_{FB}^b,R_b,m_h$) & $70\pm30$ & - & - & 5.6/12   & 0.92
\\  
\hline
SM + $S,T$ (No $A_{FB}^b,R_b$)  & $125.7\pm0.4$ & $-0.08\pm0.10$ & $0.0\pm0.08$  & 5.6/9   & 0.78
\\  
\hline
\hline
\end{tabular}
\end{center}
\caption{{\it $A_{FB}^b$ and $R_b$ due to a systematic error}:   Results of the global fits to the precision electroweak observables in Eq.~(\ref{data}), excluding those observables listed as ``No''. 
We consider 1) the SM hypothesis  and 2) the hypothesis of NP in the form of oblique parameters $S,T$. We show the Higgs mass prediction and the goodness-of-fit in terms of $\chi^2/{\rm d.o.f.}$ and $p$-value. For the second hypothesis, we also show the best-fit regions for $S, T$ parameters. }
\label{tab:fit2}
\end{table*}
%}

Next, we consider the possibility that $A_{FB}^b$ and $R_b$ are discrepant because of some unknown systematic error, perhaps related to the measurement and identification of bottom quarks. We thus exclude these observables from the fit. The results are shown in Table~\ref{tab:fit2}. The SM  describes well the rest of the data, with a $p$-value of 0.67. The direct measurement of the Higgs mass by the LHC experiments entirely determines the Higgs mass to be $m_h = 125.7$ GeV, so given the high quality of the fit,  it would seem that there is no need for NP. 

To see the tendency of the precision electroweak data (excluding $A_{FB}^b$ and $R_b$ to favor very light Higgs masses, we have also performed the fit while excluding the LHC measurement of $m_h$. One observes that the fit selects $m_h = 70 \pm 30$ GeV. One interpretation is that the indirect determination displays a slight tension with the direct determination, and one could then speculate about NP. 

For example, one way to `reconcile' a heavy Higgs with precision electroweak data is to invoke NP in the form of oblique parameters $S,T$. We have thus done a fit to the data allowing $S$ and $T$ to vary. We find that the quality of the fit is slightly improved compared to the SM case, with a $p$-value of 0.78.

\paragraph{\bf Summary}

There is no known reason to attribute the  $A_{FB}^b, R_b$ discrepancies to a systematic effect, though since both measurements rely on $b$ quark identification, it is conceivable that there is an unknown systematic effect. 
In any case, removing these discrepant observables from the global fit, 
the remaining data are in excellent agreement with the SM (quantified by a sizable $p$-value in the corresponding fit), so it is difficult to make a strong case for NP. 
There is only a modest improvement to the fit if one invokes the oblique parameters $S$ and $T$.

On the other hand, if one takes the data at face value, then there exists a preference for some NP which gives a non-universal correction to the $Z \bar b  b$ couplings. 
In our view this is the most compelling explanation of the $A_{FB}^b, R_b$ discrepancies. Below we will investigate NP models which can cause such non-universal shifts. 

\section{ Correlating $A_{FB}^b$, $R_b$ with Higgs Data}

The models that we will consider to address the discrepancies in the precision electroweak data, namely $A_{FB}^b$ and $R_b$, will also cause modifications to the properties of the Higgs boson. There are three effects that occur: 1) the coupling of the Higgs boson to bottom quarks, $y_{hbb}$ is modified, 2) new colored particles contribute to the process $h\rightarrow g g$, and 3) new charged particles contribute to the process $h \rightarrow \gamma \gamma$. These three effects can be described in a model independent fashion by the quantities, $r_\gamma$, $r_g$, $r_b$, defined as
\begin{eqnarray}
r_\gamma & = & \frac{ \Gamma( h \rightarrow \gamma \gamma) }{ \Gamma( h \rightarrow  \gamma \gamma)_{\rm SM}}, \label{rgam} \\
r_g & = &  \frac{ \Gamma( h \rightarrow gg) }{ \Gamma( h \rightarrow  gg)_{\rm SM}}, \label{rg} \\
r_b & = &  \frac{ \Gamma( h \rightarrow b\bar b) }{ \Gamma( h \rightarrow b \bar b)_{\rm SM}}. \label{rb}
\end{eqnarray}
The experimental collaborations have presented measurements of the best-fit signal strength $\mu$ of the new state in the channel $i$, defined as 
\begin{equation}
\mu =\frac{ \sigma(pp\rightarrow h\rightarrow i)}{\sigma(pp\rightarrow h\rightarrow i)_{\rm SM}}.
\end{equation} 
This dataset can then be used to constrain or test for the presence of new physics, as has been done first in Ref.~\cite{Batell:2011pz},
after the initial December 2012 announcement of an excess at 125 GeV. For recent fits done after the discovery of the 125 GeV boson, see Refs.~\cite{Hfitter}. 

In terms of these quantities, the Higgs signal strengths are predicted to be
\begin{eqnarray}
\mu_{\gamma \gamma} & \simeq & \frac{r_g r_\gamma}{1+{\rm Br}_b (r_b - 1) +  {\rm Br}_g (r_g-1) }, \label{ssgam}\\
\mu_{VV}  & \simeq & \frac{r_g}{1+{\rm Br}_b (r_b - 1) +  {\rm Br}_g (r_g-1) },  \label{ssV} \\
\mu_{b\bar b} & \simeq  & \frac{r_b}{1+{\rm Br}_b (r_b - 1) +  {\rm Br}_g (r_g-1) },  \label{ssb}
\end{eqnarray}
where ${\rm Br}_i$ denotes the SM Higgs branching ratio in the channel $i$ and the signal strengths are defined as 
\begin{equation}
\label{eq:sigstrength}
\mu_i =\frac{ \sigma(pp\rightarrow h\rightarrow i)}{\sigma(pp\rightarrow h\rightarrow i)_{SM}}.
\end{equation}
We have done a simple fit to the combined signal strength data using the $\gamma \gamma$, $ZZ$, $WW$, and $b\bar b$ channels from ATLAS~\cite{ATLAS-Higgs}, CMS~\cite{ATLAS-Higgs}, and the $\gamma \gamma$, $WW$, and $b\bar b$ channels from the Tevatron experiments ~\cite{:2012zzl}. We do not include the $\tau \bar \tau$ channel in the results presented here, but have checked that this does not qualitatively affect our conclusions. 

\begin{figure}[h!]
\begin{center}
\includegraphics[width=.45\textwidth]{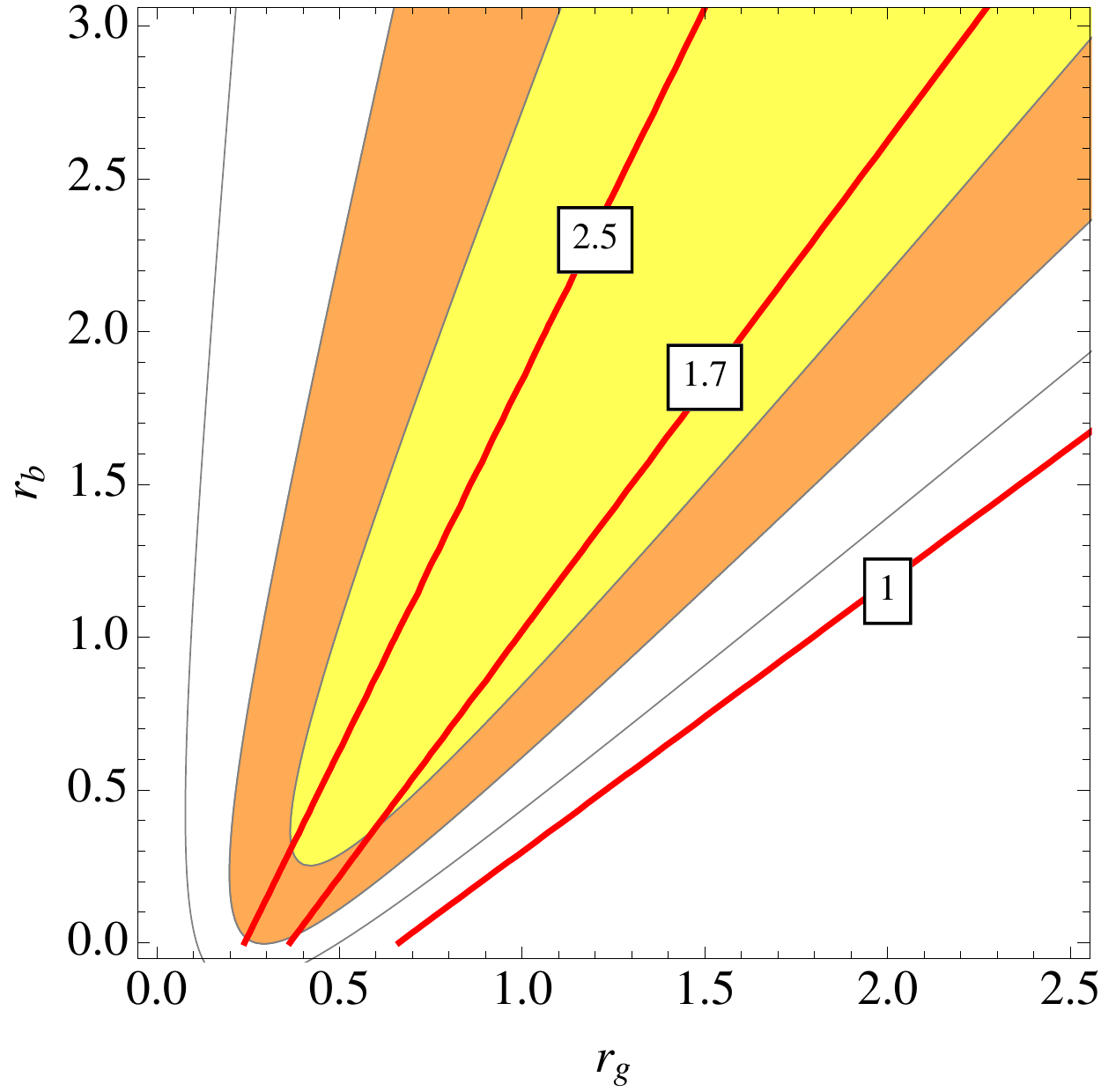}
\caption{Best-fit regions of the Higgs signal strength data $(1,2,3\sigma)$ in the $r_g-r_b$ plane. Here we have marginalized over $r_\gamma$, and contours of constant $r_\gamma$ are represented by the red lines. There is a quasi-flat direction in the $\chi^2$ function along $r_b \sim 2\, r_g$ for large value of $r_{b,g}$. }
\label{fig:rgrb}
\end{center}
\end{figure}

The best fit point yields $(r_\gamma, r_g, r_b) = (2.1,1.4,2.1)$, with a $\chi^2/{\rm d.o.f} = 6.3/8$ versus $\chi^2/{\rm d.o.f} = 11.1/11$ for the SM.
However, there is a strong degeneracy in the parameters $r_g$ and $r_b$, in that  similar fit qualities can be achieved for different combinations of these parameters, as is illustrated in Fig.~(\ref{fig:rgrb}). For example,  the points $(r_\gamma, r_g, r_b) = (2.1,0.7,0.7),~(2.1,0.8,1),~(2.1,1,1.5)$ give similar fit qualities as the best-fit point.
One can see this very easily in the regime where $r_{b,g}$ are large, in which case the signal strength predictions in Eqs.~(\ref{ssgam},\ref{ssV},\ref{ssb}) become dependent only on the ratio $r_b/r_g$ rather than each individual parameter $r_g$ and $r_b$. This degeneracy can eventually be broken by separating the diphoton data into categories sensitive to the production mechanism of the Higgs boson, in particular the VBF production mode. We have checked, however, that the degeneracy is not broken with the current dataset sensitive to VBF production due to the limited precision in these channels.  

\section{ Models for $A_{FB}^b$  and $R_b$ and Higgs physics}

We now investigate NP models that address the $A_{FB}^b$  and $R_b$ discrepancies in the precision electroweak data.
We will consider the ``Beautiful Mirrors'' scenario~\cite{Choudhury:2001hs} as our canonical example of NP that directly alters $A_{FB}^b$ and $R_b$ and simultaneously leads to interesting deviations in Higgs phenomenology. The basic mechanism for correcting $A_{FB}^b$ and $R_b$ is mixing between the bottom quark of the SM and new vector-like `mirror' quarks. This causes a modification of the coupling of the $Z$ boson to the physical bottom quark, thus affecting $A_{FB}^b$ and $R_b$. 

We start with the parameterization of the $Z \bar b b$ interaction in Eq.~(\ref{eq:Zbb}) in terms of the shifts 
$\delta g_{Lb,R b}$. According to our fits of the electroweak data in the previous section, two viable regions of parameter space exist (see Table~\ref{tab:fit2} and Fig.~\ref{fig:dg}):
\begin{eqnarray}
{\rm  I.}~~~ &  \delta g_{Rb} \sim 0.016 \pm 0.005 ,   ~~~ & \delta g_{Lb}\sim 0.001 \pm 0.001, ~~~~\label{positive}  \\
{\rm II.}~~~ &  \delta g_{Rb} \sim -0.17 \pm 0.005,  ~~~ &  \delta g_{Lb}\sim 0.001 \pm 0.001.~~~~  \label{negative} 
\end{eqnarray}
For the simplest case in which the bottom quark $b$ mixes with a single vector-like quark $B$, the shifts in the couplings are given by $\delta g_{Lb} = (t_{3}  + 1/2) s_L^2$ and $\delta g_{Rb} = t_{3} s_R^2$ ~\cite{Choudhury:2001hs}, where $t_{3}$ is the $SU(2)_L$ diagonal generator for the vector-like quark and $s_{L,R}\equiv \sin\theta_{L,R}$ are the sine of the left and right mixing angles in the $b-B$ sector. Therefore, to obtain the dominant shift in $\delta g_{Rb}$, the vector-like quarks must have nontrivial $SU(2)_L$ quantum numbers. Combined with the requirement of a charge $Q_B=-1/3$ component, there are just three possible representations. One representation, $\Psi \sim (3,3,2/3)$, allows for mixing with $b_L$ and thus leads to a larger shift in $\delta g_{Lb}$ compared to $\delta g_{Rb}$, in contrast to the pattern indicated by the data (\ref{positive},\ref{negative}). 
Thus, there are only two viable representations:
\begin{eqnarray}
\Psi^T & \sim & (T, B) \sim (3,2,1/6), \label{standard} \\
\Psi^T & \sim & (B, X) \sim (3,2,-5/6). \label{exotic}
\end{eqnarray}
We now briefly review the phenomenology of each model in its minimal implementation.

The first representation (\ref{standard}) has the same quantum numbers as the $SU(2)_L$ quark doublet of the SM, and thus contains a charge 2/3 component $T$ and charge -1/3 component $B$. Since $t_{3} = -1/2$, mixing between $b$ and $B$ leads to a negative shift in the $Z \bar b_R b_R$ coupling,  $\delta g_{Rb}=- s_R^2/2$. To resolve the $A_{FB}^b$ discrepancy, this should be fixed to $\delta g_{Rb} \sim -0.17$~(\ref{negative}), implying a sizable mixing angle $s_R \sim 0.58$.
The mixing originates from the Yukawa couplings. The relevant mass terms in the $b-B$ sector are
\begin{equation}
-{\cal L} \supset 
y_1 \bar Q H b_R +y_2 \bar \Psi_L H b_R + M \bar \Psi_L \Psi_R +{\rm h.c.}~.
\label{standardL}
\end{equation}
The required large mixing angle $\theta_R$ is obtained if the Yukawa coupling is order one, 
$Y_2 \sim 0.7 M$, with $Y_i \equiv y_i v/\sqrt{2}$, $v = 246$ GeV. A similar large mixing in the $t-T$ sector would induce a large $W \bar b_R t_R $ coupling, which is constrained from $b\rightarrow s \gamma$~\cite{Choudhury:2001hs}. Thus a large breaking of custodial symmetry is required, inducing a sizable mass splitting of the $T,B$ partners and hence a large positive NP contribution to the $T$ parameter. The electroweak data then point to a region with light vector-like quarks in the 100-200 GeV range and a heavy Higgs, far above 125 GeV.  Given the observation of the 125 GeV Higgs-like state, as well as stringent constraints on new quarks in the few hundred GeV range, this model now seems disfavored.

The second representation (\ref{exotic}) with hypercharge -5/6 contains a charge -1/3 component $B$, as well as an exotic partner $X$ with charge -4/3. In this case $t_{3}= + 1/2$, so that the shift in the coupling $\delta g_{Rb} = s_R^2/2$ is positive and is therefore fixed to the value $\delta g_{Rb} \sim 0.016$ ~(\ref{positive}) to correct $A_{FB}^b$.
This implies a smaller required mixing $s_R\sim 0.2$. Analogously to the (3,2,1/6) model, the mass terms in the $b-B$ sector that lead to such mixing are
\begin{equation}
-{\cal L} \supset 
y_1 \bar Q H b_R +y_2 \bar \Psi_L  H^\dag b_R + M \bar \Psi_L \Psi_R +{\rm h.c.}~.
\label{exoticL}
\end{equation}
In this simple implementation, there is no other fermion that can mix with the exotic quark $X$.
One may obtain the required mixing angle with a relatively small Yukawa coupling $y_2$ so that the custodial breaking is small. Thus, NP contributions to $S$ and $T$ are generically small and the model has an overall good agreement with electroweak precision data.

Models with vector-like fermions that couple to the Higgs can lead to modifications of the effective Higgs couplings; for recent studies see Refs.~\cite{VLfermion},\cite{Joglekar:2012vc},\cite{ArkaniHamed:2012kq},\cite{Reece:2012gi},\cite{Voloshin:2012tv},\cite{McKeen:2012av}
\footnote{  
See Ref.~\cite{Draper:2012xt}
 for a mechanism to alter $h\rightarrow \gamma \gamma$ without new charged states.
 }. 
 In the simple models discussed here with vector-like $B$ quarks, there is a reduction of the $h b \bar b$ coupling (by a factor $c_R^2$) and to NP contributions to $h\rightarrow gg$ and $h\rightarrow\gamma\gamma$ coming from loops of the new mirror quarks. The Higgs phenomenology for the $(3,2,1/6)$ model in Eq.~(\ref{standardL})  was studied in detail in Ref.~\cite{Morrissey:2003sc}. In particular,  for a 125 GeV Higgs the model predicts a large enhancement in the $WW$ and $ZZ$ channels, $\mu_{VV} \sim 2.4$, in conflict with the measurements for the LHC and Tevatron experiments, $\mu_{VV} \sim 1$. On the other hand, the model in Eq.~(\ref{exoticL}) leads only to small deviations from the SM, of order $10\%$ or less.

The models described by the Lagrangians (\ref{standardL}) and (\ref{exoticL}) are the minimal ones that are able to cause a sizable shift in the coupling $\delta g_{Rb}$. However, the fit to the precision data (see Fig.~\ref{fig:dg} and Eqs.~(\ref{positive},\ref{negative})), also suggests a small shift in $\delta g_{Lb}$, which cannot occur in the minimal Lagrangians. In order to bring about a small shift $\delta g_{Lb}$, Ref.~\cite{Choudhury:2001hs} added a $SU(2)_L$ singlet vector-like fermion with quantum numbers (3,1,-1/3), which has a small mixing with the SM $b_L$. The addition of such fermions can lead to additional effects in the Higgs sector beyond those in the minimal models due to a possibility of a large coupling between vector-like fermion doublet, singlet and Higgs. However, such a large coupling, by itself will lead to a sizable $T$ parameter, due to the large splitting among the components of the $SU(2)_L$ doublet fermion. This can be easily remedied in a model with an approximate custodial symmetry.
In order to exhibit the full range of Higgs phenomenology, we will now explore in detail an example of such a `Custodial' Beautiful Mirrors model.

\subsection{Custodial Model}

We consider an extension of the SM with the following vector-like fermions:
\begin{eqnarray}
\Psi^{T'}_{L,R}  &=& \left(  B', X' \right) \sim ( 3, 2, -5/6), \nonumber \\
\hat B'_{L,R}   & \sim & ( 3, 1, -1/3),  \\
\hat X'_{L,R}  & \sim &( 3, 1, -4/3).  \nonumber 
\end{eqnarray}
We will use primed fields to denote gauge eigenstates and unprimed fields to denote mass eigenstates. 
The $B'$ and $X'$ quarks have electric charges or $Q_B = -1/3$ and $Q_X = -4/3$, respectively. 

We will allow mixing of the $B$ vector-like fermions with the $b$ quark of the SM. The most general Lagrangian leading to the fermions masses is given by
\begin{eqnarray}
\label{model}
-{\cal L} & \supset &  
  M_1 \bar \Psi'_L \Psi'_R 
+M_2 \bar {\hat B}_L'  \hat B_R' +  M_3 \bar {\hat X}'_L \hat X'_R 
\nonumber \\ & + & 
y_1 \bar Q'_L H b'_R    +  y_2 \bar Q'_L H \hat B'_R  \nonumber \\
& + &  
y_3 \bar \Psi'_L \tilde H b'_R   
+ y_4 \bar \Psi'_L  \tilde H \hat B'_R   \nonumber \\ 
& + &  y_5 \bar {\hat B}_L' \tilde H^\dag \Psi'_R \nonumber \\
 & + & y_6 \bar \Psi'_L   H \hat X'_R  \nonumber \\
&+& y_7 \bar {\hat X}_L'  H^\dag \Psi'_R 
+ {\rm h.c.} ~.
\label{Lcustodial}
\end{eqnarray}
Without loss of generality, we have rotated away a possible mass term  $M \bar {\hat B}_L'  B_R'$. We note that the $\hat B'$ and $\hat X'$ fermions can be embedded in a doublet representation of a global $SU(2)_R$ symmetry, and in the limit $y_4=y_6$, $y_5 = y_7$, and $y_2 = y_3 = 0$, the Lagrangian of the mirror fermions is symmetric under the custodial symmetry $SU(2)_L\times SU(2)_R$ which protects the model from  contributions to the $T$ parameter. To correct $A_{FB}^b$ and $R_b$, we will need to take $y_2$ and $y_3$ non-zero, but still small in comparison to the vector-like mass terms, so that the custodial breaking is small.
For simplicity, we will only consider real couplings, but see Refs.~\cite{Voloshin:2012tv},\cite{McKeen:2012av} for recent studies of CP-violating phases in vector-like fermion models which modify Higgs couplings.

The Lagrangian (\ref{Lcustodial}) leads to the mass matrices in the $B$ ($b',B',\hat B'$) and $X$ ($X',\hat X'$) sectors:
\begin{eqnarray}
{\cal M}'_{B} = 
\left( 
\begin{array}{ccc}
Y_1 & 0 & Y_2 \\
Y_3 & M_1 & Y_4 \\
0 & Y_5 & M_2 
\end{array}
\right),~~~ 
{\cal M}'_X =
\left( 
\begin{array}{cc}
M_1 & Y_6  \\
Y_7 & M_3 
\end{array}
\right), ~~~
\label{LagMH}
\end{eqnarray}
where $Y_i \equiv y_i v/\sqrt{2}$. The mass matrices are 
are diagonalzied via the orthogonal transformations: 
\begin{eqnarray}
 &  {\cal M}_B = {\cal U}_B^T {\cal M}'_B {\cal W}_B  ,~~~~~~ 
&  {\cal M}_X = {\cal U}_X^T {\cal M}'_X {\cal W}_X  ,~~~~~~ 
\end{eqnarray}
Starting from the Lagrangian parameters, we can compute the mass eigenvalues in the quark sector, the rotation matrices ${\cal U}_{B,X}$, ${\cal W}_{B,X}$, and the couplings of the physical fields. 
However, as we will discuss in detail momentarily, the masses of the new quarks are quite constrained by LHC searches, and as such must be heavy. It is therefore useful to work in the regime where the vector-like mass terms $M_{1,2,3}$ are much larger than the corresponding Yukawa terms $Y_i$, and integrate out the heavy vector-like fields. We can then match to a set of effective operators that can be used to describe the corrections to the SM predictions for {\it e.g.} the $Z \bar b  b$ couplings and Higgs couplings. We now describe these steps and the approximate results in the effective theory.

\paragraph{\bf Effective theory}

We integrate out the new vector-like fermion fields at tree level, in the regime of large $M_{1,2,3}$. The procedure is as follows: we use the classical equations of motion to solve for  $\Psi$, $\hat B$, $\hat X$ in terms of the light fields $Q, b_R$, and substitute these solutions back into the Lagrangian. We then make use of the classical equations of motion for $Q, b_R$ to match to the following operator basis:
\begin{eqnarray}
\label{OP1}  {\cal O}_{Hb} & = & i ( H^\dag D_\mu H )(\bar b_R \gamma^\mu b_R  ) + {\rm h.c.}, \\  
\label{OP2} {\cal O}_{HQ}^s & = & i ( H^\dag D_\mu H )(\bar Q \gamma^\mu Q ) + {\rm h.c.}, \\
\label{OP3} {\cal O}_{HQ}^t & = & i ( H^\dag \sigma^a D_\mu H )(\bar Q \gamma^\mu \sigma^a Q ) + {\rm h.c.}, \\
\label{OP4} {\cal O}_{HY} & = & (H^\dag H) (\bar Q H b_R ) + {\rm h.c.}, \\
\label{OP5} {\cal O}_{Y} & = & (\bar Q H b_R ) + {\rm h.c.}.
\end{eqnarray}
The effective Lagrangian is given by
\begin{equation}
{\cal L} = \sum_i a_i {\cal O}_i.
\end{equation}
The first three operators above (\ref{OP1},\ref{OP2},\ref{OP3}) cause modifications to the $Z \bar b  b$ vertex, given by
\begin{equation}
\delta g_{Lb} =-  \frac{  (  a^s_{HQ}+a^t_{HQ} )  v^2 }{ 4 }    , ~~~~ 
\delta g_{Rb} = - \frac{a_{Hb} v^2 }{ 4 } . 
\end{equation}
The last two operators equations above (\ref{OP4},\ref{OP5}) affect the bottom quark mass and the $h b \bar b$ coupling:
\begin{eqnarray}
m_b &  = &- \left( \frac{a_Y  v}{\sqrt{2}} +   \frac{a_{HY}  v^3}{2 \sqrt{2} } \right), \\
y_{hbb} & = & -\left( \frac{a_Y }{\sqrt{2}} +   \frac{3 a_{HY}  v^2}{2 \sqrt{2} } \right).
\end{eqnarray}
 
After integrating out the heavy fermions and matching to the operator basis above, we obtain the following coefficients for the operators:
\begin{eqnarray}
& a_{Hb} =  \displaystyle{   - \frac{y_3^2}{2 M_1^2}  }  , ~~~~
a_{HQ}^s = \displaystyle{  -\frac{y_2^2}{4 M_2^2}}  , ~~~~
a_{HQ}^t = \displaystyle{  - \frac{y_2^2}{4 M_2^2}  },  &~~~~~~ \nonumber \\ 
& a_{Y} = -y_1, ~~~~ a_{HY} = \displaystyle{  \frac{ y_1 y_3^2}{2 M_1^2} +\frac{y_1 y_2^2}{2 M_2^2}  - \frac{y_2 y_3 y_5}{M_1 M_2}   }.&~~~~~~~
\end{eqnarray}
We thus obtain the approximate expressions for the bottom quark mass, the $hb \bar b$ coupling, and the shifts to the $Z \bar b  b$ vertex:
\begin{eqnarray}
\label{MbApprox}
& m_b = \displaystyle{ Y_1\left(  1 - \frac{Y_2^2}{2 M_2^2}    - \frac{Y_3^2}{2 M_1^2}  \right) + \frac{Y_2 Y_3 Y_5 }{M_1 M_2} },& ~~~~ \\
\label{ghbbApprox}
 & y_{hbb} = \displaystyle{  \frac{1}{v} 
\left[ 
 Y_1\left(  1 - \frac{3 Y_2^2}{2 M_2^2}  -   \frac{3 Y_3^2}{2 M_1^2}  \right) + \frac{3 Y_2 Y_3 Y_5 }{M_1 M_2}
\right] } , & ~~~~ \\
\label{dgApprox}
& \delta g_{Lb} =\displaystyle{  \frac{Y_2^2}{2 M_2^2} },  ~~~~  \delta g_{Rb} = \displaystyle{  \frac{Y_3^2}{2 M_1^2} }.&  ~~~~
\end{eqnarray}

\paragraph{\bf Corrections to $Z \bar b  b$ and $h \bar b b$ couplings}
 
As discussed above, the mirror quark representation   (3,2,-5/6) will lead to a strictly  positive shift in the coupling $\delta g_{Rb}$, and this is seen in Eq.~(\ref{dgApprox}).
Thus, we will fix the shifts $\delta g_{Lb}$ and $\delta g_{Rb}$ to their best-fit value in Eq.~(\ref{positive}) in order to bring the $A_{RB}^b$ and $R_b$ predictions into accord with the experimental values.  We then obtain from Eq.~(\ref{dgApprox})  the following relations between the Lagrangian parameters:
\begin{equation}
\label{Y2Y3}
Y_2 \simeq  \pm  0.04\, M_2, ~~~ Y_3 \simeq  \pm  0.17 \,M_1  .\ ~~~~
\end{equation}

Using Eq.~(\ref{dgApprox}), we can rewrite the approximate expression for the  bottom mass (\ref{MbApprox})and $h  \bar b b$ coupling (\ref{ghbbApprox}) in terms of the shifts $\delta g_{Lb}$, $\delta g_{Rb}$. Since $\delta g_{Lb,Rb} \ll 1$, we can express $r_b$, the ratio of the $h\rightarrow b \bar b$ partial width to its value in  the SM, defined in Eq.~(\ref{rb}) as
\begin{eqnarray}
\label{rbAP}
r_b & = &  \left(\frac{y_{hbb}}{m_b/v}\right)^2 \\
&  \simeq &  1-4(\delta g_{Rb} + \delta g_{Lb}) + 8 \sqrt{\delta g_{Rb} \delta g_{Rb}} \frac{Y_5}{m_b},  \nonumber
\end{eqnarray}
where we have taken $Y_2$, $Y_3$ to be positive. Thus, we see that only for $Y_5 \gg m_b$ can $r_b$ deviate significantly from unity.

\paragraph{\bf Low energy Higgs theorems}

Here we will apply the low energy Higgs theorem~\cite{Ellis:1975ap},~\cite{Shifman:1979eb} to derive approximate expressions for the corrections to $h \rightarrow \gamma \gamma$ and $h \rightarrow gg$. In all generality, integrating out heavy fermions $f$ charged under electromagnetism (color) yields the following effictive Lagrangian:
\begin{eqnarray}
{\cal L} & \supset & \frac{\alpha}{16 \pi v}    \left[ \sum_{f=B,X}  b_f^{EM} z_f   ~\right]  h F_{\mu\nu} F^{\mu\nu} \nonumber   \\
& +&  \frac{\alpha_s}{16 \pi v}   \left[ \sum_{f=B,X}  b_f^{c} z_f ~ \right] h G^a_{\mu\nu} G^{\mu\nu a},
\end{eqnarray}
where the quantities $z_f$ are given by 
\begin{equation}\label{eq:zf}
z_f \equiv  \frac{\partial}{\partial \log v} \left( \sum_{i} \log m_{f,i}^2(v)  \right),
\end{equation}
and the sum is over all heavy fermions.
The beta function coefficients are in general $b_f^{EM} = (4/3)N_c Q_f^2$, $b_f^c = (4/3)C(r)$, where $N_c$, $Q_f$ and $C(r)$ are the number of colors, the electric charge and the Casimir of the particular representation of the fermion $f$. For the system at hand we have
\begin{equation}
b_B^{EM} = 4/9, ~~~~ b_X^{EM} = 64/9, ~~~~ b_B^{c} = b_X^c = 2/3.
 \end{equation}

Given the effective Lagrangian, we can derive the ratios $r_{\gamma,g}$ as
\begin{eqnarray}
r_\gamma 
= \bigg\vert 1- \frac{1}{2} \frac{ b_B^{EM} z_B + b_X^{EM} z_X }{A_\gamma^{SM}}  \bigg\vert^2,  \nonumber \\
r_g 
= \bigg\vert 1+  \frac{ b_B^c z_B +b_X^c z_X }{A_g^{SM}}  \bigg\vert^2. ~~~~~~
\end{eqnarray}
with $A_g^{SM} \simeq 1.3$ and $A_\gamma^{SM} \simeq 6.6$.

To evaluate the quantities $z_f$, it is convenient to rewrite the sum in terms of the determinant of the mass matrix, so that it can be evaluated in any basis. For the bottom sector, however, we cannot straightforwardly do this since the mass matrix contains a light eigenvalue. Let us then derive a simple modification which allows us to apply the low energy theorem. We can rewrite Eq.(\ref{eq:zf}) as
\begin{eqnarray}
z_B &= &\frac{\partial}{\partial \log v} \sum_{i=2}^3 \log m_{B,i}^2(v)    \nonumber \\
 & = & \frac{\partial}{\partial \log v} \log \det {\cal M}_B {\cal M}_B^\dag  -  
 \frac{2v}{m_b(v) } \frac{\partial}{ \partial v} m_b(v), ~~~~~~
 \label{LETb}
\end{eqnarray}
where the sum on the first line is only over the heavy eigenvalues of the mass matrix $\cal M'$ defined in (\ref{LagMH}) and is
the usual piece associated to the use of the low energy theorem.  Since the determinant includes a light bottom quark contribution, we must subtract this contribution, which is accounted for by the second term above. Using the approximate expression for $m_b$ in Eq.~(\ref{MbApprox}), 
we can write the leading terms of order $1/M^2$:
\begin{eqnarray}
 z_B  & \simeq & 
\frac{ 2 Y_2^2}{M_2^2} + \frac{2 Y_3^2}{M_1^2} -\frac{ 4 Y_4 Y_5}{M_1 M_2} \nonumber \\
& \simeq &  -\frac{ 4 Y_4 Y_5}{M_1 M_2}.
\end{eqnarray}
Note that the last expression is valid in the regime $Y_{4,5} \gg  Y_{1,2}$.  In any case, since the first two terms above are directly proportional to the small shifts in the $Z \bar b  b$ couplings  (\ref{dgApprox}), they will not cause large deviations 
in the $r_{g,\gamma}$.

For the $X$ sector, we can straightforwardly convert the sum to a determinant. The result is 
\begin{eqnarray}
z_X & = &   \frac{\partial}{\partial \log v} \log \rm det\, {\cal M}_X   {\cal M}^\dag_X  \nonumber  \\
& \simeq &-\frac{ 4 Y_6 Y_7}{M_1 M_3 }.
\end{eqnarray}

Note that, due to its large electric charge of $-4/3$ (leading to $b_X^{EM}=64/9$), the $X$ fermion contribution will be the dominant one in $h\rightarrow \gamma\gamma$, and correspondingly $Y_{6,7}$ will mainly affect $r_\gamma$. On the other hand, the $B$ and $X$ contributions to $r_g$ will be comparable.  

\paragraph{\bf Oblique corrections}

The new mirror fermions in our model can give sizable contributions to the oblique parameters $S$ and $T$, especially if they have large Yukawa couplings to the Higgs field.
Given that we are invoking NP shifts to the couplings $\delta g_{Lb,Rb}$ to correct $A_{FB}^b$ and $R_b$, the allowed regions for $S$ and $T$ will be shifted somewhat compared to the usual regions derived for the SM $Z \bar b b$ couplings. 

We have thus performed a fit to the electroweak data allowing for the presence of coupling shifts  $\delta g_{Lb,Rb}$ and oblique parameters $S$ and $T$. For the $Z \bar b b $ coupling shifts, the fit selects regions essentially identical to the one displayed in Fig.~(\ref{fig:dg}), while for the oblique parameters
the preferred regions are
\begin{equation}
S = -0.02 \pm 0.09, ~~~~ T = 0.03 \pm 0.08 , 
\label{ST}
\end{equation}
with a correlation coefficient of $0.90$. Since the motivation for introducing the mirror fermions is to improve the description of the precision electroweak data, we will restrict the parameter space to the region in Eq.~(\ref{ST}) when we present our numerical results below. 

For the model under consideration, the oblique corrections are most easily computed directly in the mass basis. Our procedure will thus be to diagonalize the system (\ref{model}), obtain the physical couplings of mirror fermions to gauge bosons, and compute the vacuum polarization functions of the gauge bosons. Since the physical couplings of the top and bottom quarks to the electroweak gauge bosons are shifted in this model from the SM, we will include in our computations the contributions of those particles and at the end subtract off the SM values. 

Here we present the general expressions for the vacuum polarization functions. Given two fermions $\psi_i,\psi_j$ interacting with a vector boson $V_\mu^A$ through the Lagrangian
\begin{equation}
{\cal L} \supset   V^A_{\mu} \bar \psi_i \gamma^{\mu}(L_A^{ij}P_L   + R_A^{ij}P_R) \psi_{j},
\end{equation}
the contribution to the vacuum polarization function  $\Pi_{AB}^{ij}(p^2) $ 
arising from the one-loop exchange of these two fermions is given by:
\begin{eqnarray}
\Pi_{AB}^{ij}& = & \frac{N_c}{16 \pi^2}     \left[   
X_{AB}^{ij} F(p, m_i, m_j)+ Y_{AB}^{ij}G (p, m_i, m_j) \right], \nonumber \\ &&
\end{eqnarray}
where we have defined
\begin{eqnarray}
X_{AB}^{ij} & \equiv & L_A^{ij} L_B^{ji}+R_A^{ij} R_B^{ji}, \nonumber   \\
Y_{AB}^{ij} & \equiv & m_i m_j (L_A^{ij} R_B^{ji}+R_A^{ij} L_B^{ji}), \nonumber \\
F(p,m_i,m_j) & \equiv & 4 B_{22}(p,m_i,m_j)+ 2 p^2 (B_{21}(p,m_i,m_j) \nonumber \\
&-& B_1(p,m_i,m_j)) - m_i^2 - m_j^2 + \frac{p^2}{3}, \nonumber \\
G(p,m_i,m_j) & \equiv & -2 B_{0}(p,m_i,m_j),
\end{eqnarray}
with $B_i(p,m_i,m_j)$ are the Passarino-Veltmann functions~\cite{Passarino:1978jh}. With the vacuum polarization functions in hand, we can straightforwardly compute $S,T$ according to the definitions in {\it e.g.} \cite{Maksymyk:1993zm}.

\paragraph{\bf Collider bounds and prospects}

The collider phenomenology of this model is very similar to the `minimal' model with the $(3,2,-5/6)$ representation discussed above, and has been studied in detail in Ref.~\cite{Kumar:2010vx}. For $m_{B,X}\lesssim700$ GeV, QCD pair production has the largest rate, while for larger masses electroweak single production of mirror quarks dominates.

As recently summarized in~\cite{Okada:2012gy}, numerous searches for heavy vector-like quarks have been performed both at the Tevatron and LHC, though no evidence for the existence of exotic quarks has been observed. Here we focus on the constraints and signatures relevant for our model. 
 
The exotic bottom-type quarks $B$ arising in our model will be pair-produced at hadron colliders via
their QCD interactions. Since they only mix with the third generation down-quark of the SM, their decay modes are $B\rightarrow tW,\,bZ,\,bh$. 

The first decay mode $B\rightarrow tW$ is the most constrained by the experimental searches for vector-like $b$-quarks. Searches have been performed with final states containing only one~\cite{Aaltonen:2011vr,ATLAS:2012aw} (two~\cite{Aaltonen:2009nr,Aad:2012bb} or at least two~\cite{Chatrchyan:2012yea,Atlas2012}) $W$ bosons decaying leptonically. The strongest constraint comes from the ATLAS analysis~\cite{Atlas2012} based on 4.7 fb$^{-1}$ 7 TeV LHC data. This analysis requires final states containing at least two isolated leptons of the same charge and at least two jets, including at least one b-tagged jet: vector-like $b$-quarks decaying $100\%$ to $tW$ with mass below 670 GeV are now excluded at $95\%$ C.L. .

Furthermore, several Tevatron and LHC searches give bounds on the mass of vector-like $b$-quarks decaying $100\%$ 
to $bZ$. An early CDF analysis based on 1.06 fb$^{-1}$~\cite{Aaltonen:2007je} of data bounds the mass to be greater than 268 GeV. This bound has recently been improved by the ATLAS~\cite{:2012ak} and CMS collaborations~\cite{CMSBbZ}, using respectively 2 and 4.9 fb$^{-1}$ data at the 7 TeV LHC. Vector-like $b$-quarks decaying $100\%$ to $bZ$ with mass below 550 GeV are now excluded at $95\%$ C.L.

Finally $B$-quarks decaying to $bh$, giving 6$b$ final states are the most difficult to search for. The ATLAS and CMS collaborations are performing extensive searches for stops and gluinos. Several analysis, based on 7 and 8 TeV LHC data, are requiring final states containing multiple jets and at least 1,2,3 $b$-jets~\cite{CMS839,CMS7498,:2012pq,:2012si,:2012rg}. This searches can be recast to set bounds on the mass of vector-like $B$-quarks decaying mainly to $bh$. However, to optimize stop and gluino searches, the CMS and ATLAS analysis impose a strong requirement on the amount of missing energy (at least 100-150 GeV), which supresses the sensitivity of these searches to $B$-quarks decaying to $bh$. The constraints on the $bh$ decay mode are thus weak in comparison to the $tW$, $bZ$ modes, thus warranting a dedicated 6$b$ search.

The additional exotic $X$ quarks arising in our model will also be produced copiously in pairs and will decay to $ b W^-$ and $B W^-$. Tevatron and LHC $t^\prime$ searches thus set stringent bounds on the mass of these vector-like fermions decaying to $bW$. In particular, searches have been performed with one~\cite{Aaltonen:2011tq,Aad:2012xc,CMStprime} and two~\cite{CMS:2012ab} leptonic $W$ bosons. The CMS analysis based on 4.6-4.7 fb$^{-1}$ 7 TeV LHC data~\cite{CMStprime} sets the most stringent bound: $M_X\geq 560$ GeV. 

The longer decay chains $X\rightarrow  B W\rightarrow (tW,\,bZ,\,bh)W$ are instead more weakly constrained by present Tevatron and LHC analysis.  
However, it is quite difficult in our model to arrange for the decay $X\rightarrow  B W$. As we will see below, the Higgs signal strength data prefer Yukawa couplings to the $X$ sector that are larger than those to the $B$ sector, typically causing the lightest mirror quark to be $X$ rather than $B$. One could arrange a lighter $X$ by making $M_3 \ll M_2$, but at the expense of large custodial symmetry breaking. Thus in practice, the exotic $X$ quark in our model decays via $X \rightarrow b W$ with a $100\%$ branching ratio and the constraint from $t^\prime$ searches, $M_X>560$ GeV is quite robust. 

Given that $X$ quarks must be heavier than 560 GeV and assuming $M_X < M_B$, as indirectly suggested by Higgs signal strength data and custodial symmetry, 
in practice the bounds on $M_B$ from vector like $b$-quark searches are automatically circumvented in our model. In particular the bound $M_B>670$ GeV obtained by CMS for $b$-quarks decaying $100\%$ in $t W$ is satisfied since the $B$ quarks typically have ${\cal O}(1)$ branching ratios to $bZ$ and $bh$, weakening the limit. 

In terms of future prospects, we estimate that with the full 8 TeV data set, the bound on $X$ coming from pair production can be extended to roughly $m_X\gtrsim 700$ GeV. To probe higher masses, it will be important to focus on single production of mirror quarks.  We expect that the $\sqrt{s}=14$ TeV LHC will be able to probe mirror quarks masses into the $\sim 2$ TeV range; see~\cite{Kumar:2010vx} for a detailed study.

Finally, we mention that besides the direct production of vector-like quarks,  another possible novel signature is enhanced double Higgs production, which occurs due to the loop-level contribution of the heavy quarks to $gg\rightarrow hh$. See, for example, Refs.~\cite{dihiggs} for dedicated studies of this signature.

\paragraph{\bf Vacuum Stability}

The Yukawa couplings of the mirror fermions to the Higgs boson yield negative contributions to the $\beta$ function of the Higgs quartic coupling. For ${\cal O}(1)$ Yukawa couplings, this can cause the Higgs quartic to run negative at a low scale, potentially leading to a vacuum instability problem. This point has been emphasized recently~\cite{Joglekar:2012vc},\cite{ArkaniHamed:2012kq},\cite{Reece:2012gi} in the context of the enhanced the di-photon signal.
In the model under consideration this is also an issue as can be seen by examining the $\beta$ function for the Higgs quartic $\lambda$:
\begin{eqnarray}
16 \pi^2 \beta_{\lambda} &  \simeq & 24 \lambda^2 +
 12 \lambda (y_t^2+ y_4^2+ y_5^2+ y_6^2+ y_7^2 ) \nonumber  \\
 && - 6 (y_t^4 +y_4^4+ y_5^4+ y_6^4+ y_7^4 ). ~
\end{eqnarray}
This expression is valid in the limit of $y_{1,2,3} \ll y_t$ and of small gauge couplings. On the one hand, ${\cal O}(1)$ values for the new coupligs $y_{4,5,6,7}$ can cause desirable modifications to Higgs physics, but clearly the larger these couplings are the faster the Higgs quartic will run negative. As an example, if we take the values $y_4=y_5 = 0$ and $y_6 = y_7 = 1$ and a threshold at the vector-like mass scale of $M = 800$ GeV, we find that the Higgs quartic vanishes at the scale 2 TeV, while it can be metastable up to scales of order 3 TeV, where we have used the results of~\cite{Isidori:2001bm} to estimate the metastability bound.

Thus, the model with these parameters requires a UV completion at low scale.  One possible completion would be supersymmetric version of the model, since  the scalar superpartners of the mirror quarks can provide an equal and opposite contribution to the beta functions.  This direction, however, lies outside the scope of this paper.

\paragraph{\bf Numerical Results}

We now explore the parameter space of the model (\ref{model}), finding regions which resolve $A_{FB}^b$ and $R_b$ discrepancies, give small contributions to the oblique parameters $S$ and $T$, are consistent with direct searches for vector-like quarks at Tevatron and LHC, and, finally, provide a good description of the Higgs signal strength data. 

Since there are many new parameters in the model, we will now make several physically motivated assumptions in order to reduce the parameter space: 1) We fix the couplings $Y_2$ and $Y_3$ according to Eq.~(\ref{Y2Y3}) in order to cause the shift in the $Zb \bar b$ couplings to their central values in Eq.~(\ref{positive}), which bring the $A_{FB}^b$ and $R_b$  predictions into agreement with their measured values. 2) We fix $Y_1$ by the requirement of obtaining the correct $b$ quark mass via Eq.~(\ref{MbApprox})
3) We fix $M_2 = M_3$ as motivated by custodial symmetry.  This still leaves 5 parameters that describe the model, namely, $Y_4,Y_5,Y_6,Y_7, M_1, M_2$. For simplicity, we will further fix $Y_4 = Y_5  \equiv Y_B$ and  $Y_6 = Y_7 = Y_X$,  describing  common Yukawa couplings in the mirror $B$-quark and $X$-quark  sectors, respectively. Finally, we will also assume a common vector-like quark mass scale $M_1 = M_2 = M_3 \equiv M$. With these simplifications there are  3 parameters: $Y_B, Y_X$, and $M$.  

The most robust constraint on the model comes from the $t^\prime$ searches at the LHC discussed above, which restrict $M_X > 560$ GeV. With the simplifying assumptions above, the lightest $X$ mass is given by
\begin{equation}
M_X = M - Y_X,
\end{equation}
so that, {\it e.g.} for a given vector-like mass scale $M$, the $X$ sector Yukawa coupling must be less than some maximum value. 

\begin{figure}[h!]
\begin{center}
\includegraphics[width=.45\textwidth]{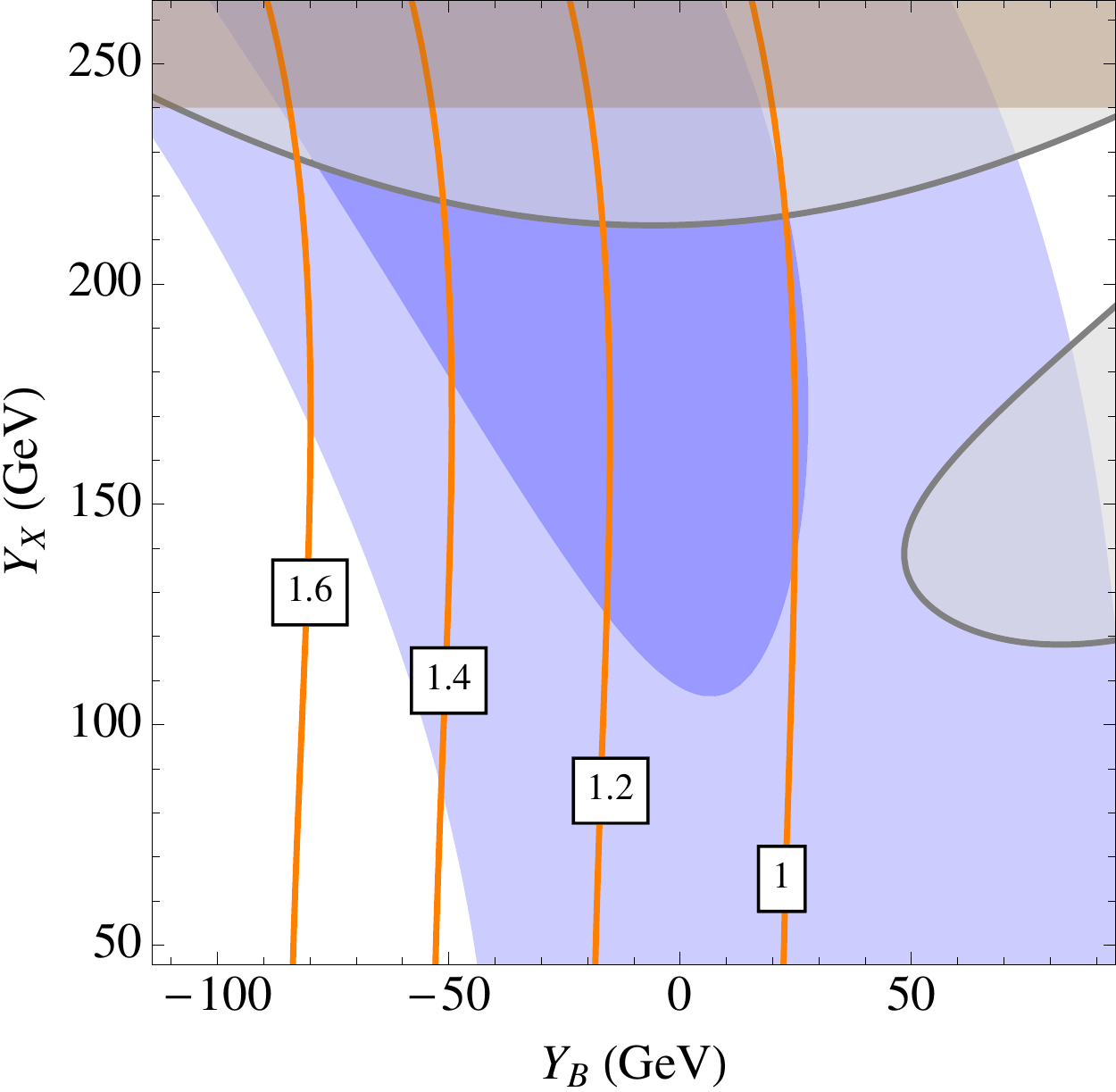}
\caption{ Preferred regions of parameter space in the $Y_B - Y_X$ plane, with the common vector-like mass fixed to $M=800$ GeV. The dark (light) blue area represents the Higgs signal strength $1\sigma$ ($2\sigma$) preferred region. The gray region predicts oblique parameters $S,T$ outside the $1\sigma$ preferred region.  The brown shaded region is excluded by $t^\prime$ searches. We also show in orange the contours of constant signal strength in the diphoton channel $\mu_{\gamma\gamma}$. }
\label{fig:result1}
\end{center}
\end{figure}

We first fix the vector-like mass scale to $M=800$ GeV and show in Fig.~\ref{fig:result1} the preferred regions of parameter space in the $Y_B - Y_X$ plane.  The fit to the Higgs signal strength selects a region  $ -100~{\rm GeV}  \lesssim Y_B \lesssim 20~{\rm GeV}$ and $|Y_X| \gtrsim 100$ GeV, shown in dark blue in Fig.~\ref{fig:result1}.  In this region, the signal strength in the diphoton channel is enhanced, $1 \lesssim \mu_{\gamma\gamma} \lesssim 1.6$, as illustrated by the constant $\mu_{\gamma\gamma}$ contours in orange in Fig.~\ref{fig:result1}. This enhancement in $\mu_{\gamma\gamma}$ is a result of 1) the loop contribution of the charged 
$-4/3$ particle $X$ to $h\rightarrow \gamma \gamma$ causing $r_\gamma \gtrsim 1$, and 2) a suppressed coupling of the physical $b$-quark to the Higgs due to mixing, which causes $r_b \lesssim 1$ and enhances the branching ratio of $h \rightarrow  \gamma\gamma$. The brown shaded region in  Fig.~\ref{fig:result1} corresponds to $M_X > 560$ GeV and is thus excluded by $t^\prime$ searches. Finally, the parameters in the gray shaded regions lead to oblique contributions which are in tension with the preferred values of our fit, given in Eq.~(\ref{ST}).

\begin{figure}[h!]
\begin{center}
\includegraphics[width=.45\textwidth]{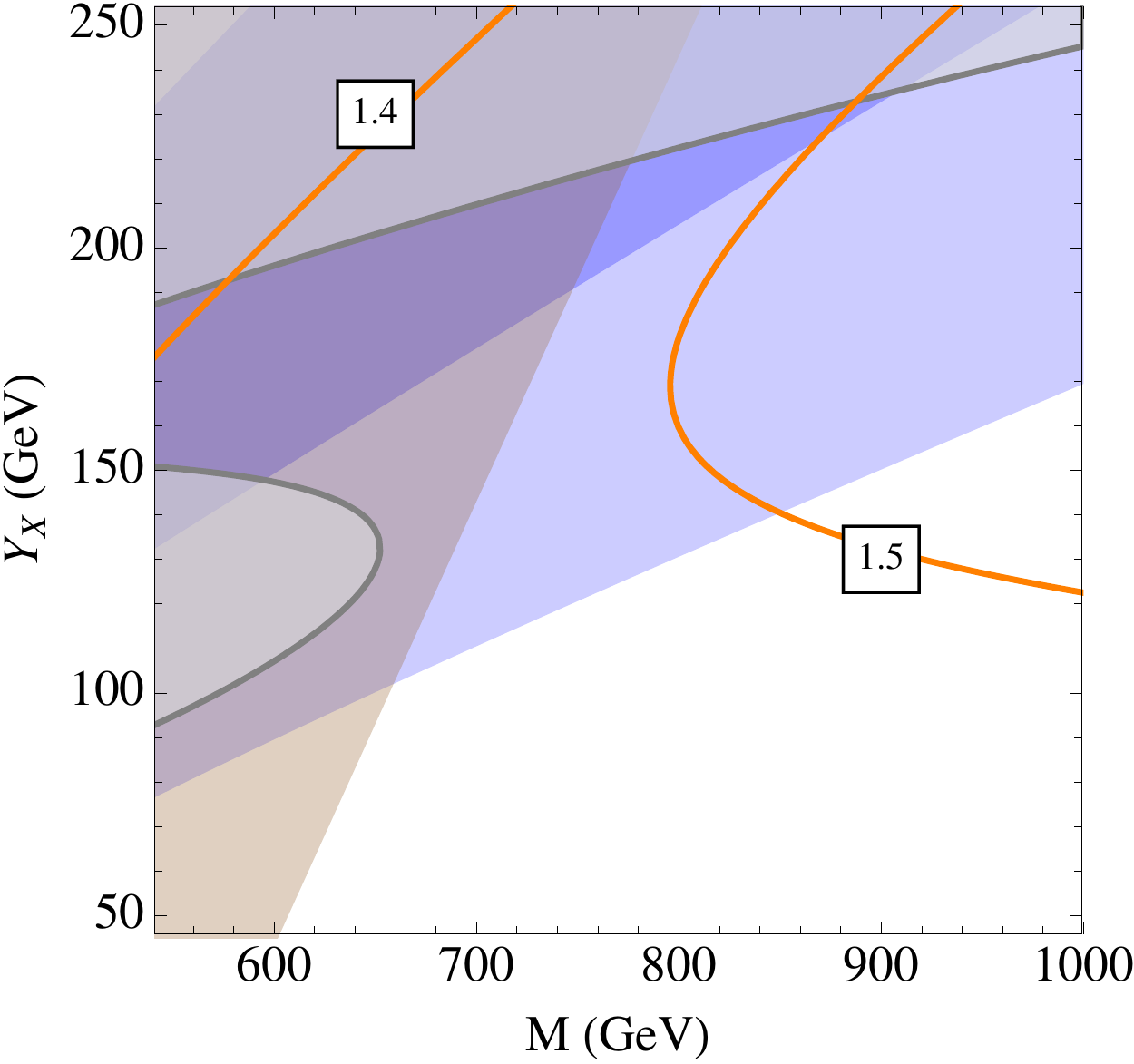}
\caption{Preferred regions of parameter space in the $M - Y_X$ plane, with the common $B$-sector Yukawa fixed to $Y_B =  - 65$ GeV. The dark (light) blue area represents the Higgs signal strength $1\sigma$ ($2\sigma$) preferred region. The gray region predict oblique parameters $S,T$ outside the $1\sigma$ preferred region.  The brown shaded region is excluded by $t^\prime$ searches. We also show in orange contours of constant signal strength in the diphoton channel. $\mu_{\gamma\gamma}$ }
\label{fig:result2}
\end{center}
\end{figure}

Next, we fix the common $B$-sector Yukawa coupling to the value $Y_B = -65$ GeV and show in Fig.~\ref{fig:result2} the preferred regions of parameter space in the $M-Y_X$ plane. 
The regions are as described above for Fig.~\ref{fig:result1}. We see that, for this value of $Y_B$, one obtains an enhancement of $\mu_{\gamma \gamma}\sim 1.5$, though only a small region of parameter space is allowed. Decreasing $Y_B$, larger regions of parameter space open up, although the enhancement in $\mu_{\gamma \gamma}$ is less. This is consistent with the regions displayed in Fig.~(\ref{fig:result1}).

For these results, we have taken $Y_2$ and $Y_3$ to be positive, so that the Higgs data selects out regions of negative $Y_B$ which suppress $r_b$ and thus enhance $\mu_{\gamma\gamma}$. The dependence of $r_b$ on the Yukawa couplings can be seen in in Eqs.~(\ref{MbApprox},\ref{ghbbApprox},\ref{rbAP}). We note that there are several physically equivalent regions related by different sign choices of the Yukawa couplings.

We concluded that the model can provide a good description to the Higgs signal strength data, and in particular an enhancement in the diphoton channel $\mu_{\gamma\gamma} \sim 1.6$, while simultaneously resolving the two discrepant precision observables $A_{FB}^b$ and $R_b$.

\section{Conclusions}

In this paper we have reexamined the precision electroweak data following the discovery of a Higgs-like state at the LHC. We have paid special attention to $A_{FB}^b$ and $R_b$, which currently exhibit tension with the SM at the level of $2.4\sigma$ and $2.3\sigma$, respectively . While the $A_{FB}^b$ discrepancy is well-known, the tension in $R_b$ is a recent development due to a state-of-the-art calculation including two-loop electroweak corrections~\cite{Freitas:2012sy}. 

One logical possibility is that these discrepancies are the result of an unknown systematic error, although no candidate for such an effect is known. Nevertheless, we have shown that upon exclusion of these observables from the global electroweak fit, the remaining data are well described by the SM, and the fit prefers a Higgs mass of 125.7 GeV due to the precise LHC measurement. While no strong argument for new physics can be made in this case, a marginal improvement in the fit is possible if NP causes oblique corrections. 

On the other hand, if one believes the measurements are correct, then the global fit to the SM is of low quality. New physics in the form of non-universal shifts to the $Z \bar b b$ vertex can dramatically improve the global fit and bring the predictions for $A_{FB}^b$ and $R_b$ into agreement with the measurements.

It is reasonable to expect that such new physics will lead also to modifications to the properties of the Higgs boson, and this interplay has been the primary focus of our work. In particular, we have investigated the Beautiful Mirrors scenario, which contains new vector-like quarks that mix with the bottom quark of the SM. This scenario generically predicts modifications to the Higgs boson couplings to $b$ quarks, gluons, and photons. We have derived in a model-independent fashion the shifts preferred by the Higgs signal strength data from the LHC and Tevatron experiments.  

In the simplest models in this framework, the new mirror quarks in this scenario can have large couplings to the Higgs boson, but only at the expense of large custodial symmetry breaking which could reintroduce a tension with the electroweak data. With this motivation, we have proposed a `custodial' version of the model which can protect against large contributions to the $T$ parameter. We have confronted this model with precision electroweak data, Higgs signal strength data, and collider searches, deriving regions of parameter space which are in agreement with all experimental results.  In particular, the model predicts an enhancement of the diphoton signal strength 
between $1\lesssim \mu_{\gamma\gamma} \lesssim 1.6$. If the current trend in the diphoton channel continues, then the scenario is testable in the immediate future, as it predicts light mirror quarks around the TeV scale which will be probed by the LHC experiments.

\subsubsection*{Acknowledgements}

We thank Carlos Wagner for helpful discussions. S.G.
and L.T.W thank the Aspen Center of Physics where part of
this work was completed. 
Work at ANL is supported in part by the U.S. Department of Energy (DOE), Div. of HEP, Contract DE-AC02-06CH11357.
L.T.W. and B.B. are supported by the NSF under grant PHY-0756966 and the DOE Early Career Award under grant
DE-SC0003930. B.B. was supported in part by the DOE under under Task TeV of contract DE-FG02-96ER40956 during the New Physics in Heavy Flavor in Hadron Colliders at the University of Washington.

\end{document}